\documentclass[twocolumn,showpacs,preprintnumbers,amsmath,amssymb]{revtex4}
\usepackage{graphicx}
\usepackage{dcolumn}
\usepackage{bm}
\begin{document}
\title{Braid Group, Gauge Invariance and Topological Order}
\author{Masatoshi Sato
}%
\email{msato@issp.u-tokyo.ac.jp}
\author{Mahito Kohmoto}
\affiliation{%
Institute for Solid State Physics,
The University of Tokyo,
Kashiwanoha 5-1-5, Kashiwa, Chiba, 277-8581, Japan
}%
\author{Yong-Shi Wu}
\affiliation{
Department of Physics, University of Utah, Salt Lake City, Utah 84112
}%
\date{\today}
\begin{abstract}
Topological order in two-dimensional systems is studied 
by combining the braid group formalism with a gauge 
invariance analysis. We show that flux insertions (or 
large gauge transformations) pertinent to the toroidal 
topology induce automorphisms of the braid group, giving 
rise to a unified algebraic structure that characterizes 
the ground-state subspace and fractionally charged, anyonic 
quasiparticles. Minimal ground state degeneracy is derived 
without assuming any relation between quasiparticle charge 
and statistics. We also point out that noncommutativity 
between large gauge transformations is essential for the 
topological order in the fractional quantum Hall effect.  
\end{abstract}

\maketitle

In contrast to the expectation from classical thermodynamics,
physics at absolute zero temperature is very rich because of novel 
effects of quantum fluctuations. In recent years it has 
become increasingly clear that in a wide class of 
two-dimensional strongly correlated many-body systems,
transition driven by a non-thermal parameter may occur 
at zero temperature to a novel phase which can not be 
described by usual spontaneous symmetry breaking and 
order parameters. The characteristic signature of the 
novel phase is a finite ground state degeneracy that 
depends on the topology of the system; accompanying with 
it are charge fractionalization (with respect to that of 
the constituent particles) and/or fractional statistics
of the quasiparticles. The first known example is the 
Laughlin state \cite{Laughlin} for the fractional quantum 
Hall (FQH) effect, with electron filling factor $\nu=1/n$ 
with $n$ odd. Soon after, it was realized that in this 
phase the ground state is $n$-fold degenerate on a 
cylinder \cite{TW} or on a torus \cite{WN}, while is 
known nondegenerate on a sphere. Actually it is the 
ground state degeneracy that is responsible for the 
fractional quantization of Hall conductance \cite{TW,NTW} 
and dictates the fractional charge $e*=e/n$ \cite{WHK} 
and the anyon statistics $\theta=\pi/n$ \cite{WN,Einarsson} 
of the quasiparticles. This type of new order is dubbed 
as topological order \cite{Wen}. In recent years more 
systems, including bosonic ones or at zero magnetic 
field, are identified to possess topological order 
\cite{Wen4,WZ2,RS91,Sent00,Moes01,Misg02,Bal02,Motr02,Free04,DST}. 
 
In the study of topological order, a central issue is how 
to characterize or classify topological orders. Previously
there has been the idea \cite{TW,WHK} that the topology 
dependent ground state degeneracy seems to be dictated by 
an (emergent) discrete symmetry. But the latter was never 
identified explicitly, except being $Z_n$ for the Laughlin 
states. Another important issue is how to understand the 
relationship between ground state degeneracy and charge 
fractionalization and/or quasiparticle statistics. A 
puzzling fact is that different patterns have 
appeared in investigations of various systems. For example, 
it was concluded \cite{WN} for the FQH systems that on a 
surface with non-zero genus $g$, the appearance of a 
fractional $\theta=\pi m/n$ statistics, with $m$ and $n$ 
co-primes, requires $n^g$-fold degenerate ground states, 
confirmed by the braid group analysis \cite{Einarsson,HKW} 
and by the effective Chern-Simons theory as well 
\cite{Wen2,WDF}. On the other hand, in a recent paper \cite{OS} 
it was shown, by using a gauge invariance argument \cite{WHK}, 
that charge fractionalization with $e^{*}=ep/q$, with $p$ 
and $q$ co-primes, requires a ground state degeneracy $q^{2g}$ 
if the quasiparticles are ordinary bosons and fermions, 
while the FQH ground state degeneracy is known to be only 
$q^g$-fold \cite{WN,HKW}.  

In this letter, we start with a reexamination of the 
interplay between charge fractionalization and quasiparticle 
anyon statistics, if they coexist, in constraining ground 
state degeneracy. This has been studied only for the Laughlin 
states and their variants \cite{WN,HKW}, where fractional 
charge and anyon statistics are known to be closely related 
to each other \cite{ASW}. Below we shall derive minimal 
ground state degeneracy without assuming any relation 
between quasiparticle charge and statistics. A bonus of 
our reexamination is the identification of the discrete 
topological symmetry algebra that underlies the ground 
state degeneracy, which can be used to classify topological 
orders that support Abelian anyonic quasiparticle excitations.  
 
\begin{figure}[h]
\begin{center}
\includegraphics[width=7cm]{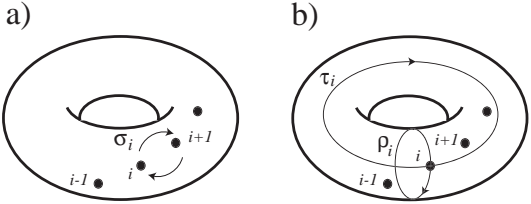}
\caption{a) An exchange of quasiparticles; 
b) Translations along non-shrinkable loops on 
the torus.}
\label{fig:braidloop}
\end{center}
\end{figure}

We will start with the braid group formalism 
\cite{Wu,Einarsson,HKW2} for fractional statistics. Consider 
for $N$ quasiparticles in a toroidal system with size 
$L_x\times L_y$. The braid group generators \cite{Birman}
consist of $\sigma_i$ $(i=1,\cdots, N-1)$, which exchanges the 
$i$th and $(i+1)$th particles clockwise without 
enclosing any other quasiparticle, and of $\tau_i$ and 
$\rho_i$ $(i=1,\cdots, N)$, which represent moving the 
$i$-th quasiparticle along a loop on the torus in 
$x$- and $y$-direction, respectively. (See 
Fig.\ref{fig:braidloop}.) Define operators 
$A_{i,j}$ and $C_{i,j}$ as 
\begin{eqnarray}
A_{j,i}=\tau_j^{-1}\rho_i\tau_j\rho_i^{-1},
\quad
C_{j,i}=\rho_j^{-1}\tau_i\rho_j\tau_i^{-1},
\end{eqnarray}
where $1\le i<j\le N$. The exchange operators 
$\sigma_i$ satisfy the 
following relations,
\begin{eqnarray}
&&\sigma_k\sigma_l=\sigma_l\sigma_k,
\quad 1\le k\le N-3, \quad |l-k|\ge 2,
\nonumber\\ 
&&\sigma_k\sigma_{k+1}\sigma_k
=\sigma_{k+1}\sigma_k\sigma_{k+1}, 
\quad 1\le k \le N-2,  
\nonumber\\
&& \tau_{i+1}=\sigma_{i}^{-1} \tau_i \sigma_i^{-1},
\quad 
\rho_{i+1}=\sigma_i \rho_i \sigma_i,
\nonumber\\
&& \tau_1\sigma_j=\sigma_j\tau_1, 
\quad 
\rho_1\sigma_j=\sigma_j\rho_1, \quad 
\sigma_i^2=A_{i+1,i}, 
\label{eq:braid1}
\end{eqnarray}
where $1\le i\le N-1$ and $2\le j\le N-1$.
And $\tau_i$'s and $\rho_i$'s satisfy
\begin{eqnarray}
&&A_{m,l}\tau_k=\tau_k A_{m,l},
\quad
A_{m,l}\rho_k=\rho_k A_{m,l}
\nonumber\\
&&\tau_i \tau_j=\tau_j \tau_i, 
\quad
\rho_i\rho_j=\rho_j\rho_i,
\nonumber\\
&&C_{j,i}=(\tau_i\tau_j)A_{j,i}^{-1}(\tau_j^{-1}\tau_i^{-1}),
\quad
A_{j,i}=(\rho_i\rho_j)C_{j,i}^{-1}(\rho_j^{-1}\rho_i^{-1}),
\nonumber\\
&&C_{j,i}=(A_{j,j-1}^{-1}\cdots A_{j,i+1}^{-1})A_{j,i}^{-1}
(A_{j,i+1}\cdots A_{j,j-1}),
\nonumber\\
&&\tau_1\rho_1\tau_1^{-1}\rho_1^{-1}
=A_{2,1}A_{3,1}\cdots A_{N-1,1}A_{N,1},
\label{eq:braid2}
\end{eqnarray}
where $1\le k <l<m\le N$ and $1\le i<j\le N$.

Let us now assume that the quasiparticles have a fractional 
charge $e^{*}=(p/q)e$, where $p$ and $q$ are mutually prime 
integers, and consider an adiabatic insertion of flux 
$2\pi/e$ through one of the holes of the torus. If the 
adiabatic flux insertion induces an infinitesimal electric 
field in $x$-direction, the process can be realized by 
a large gauge transformation $U_x$, in which the 
$x$-component of the gauge field changes from $A_{x}=0$ 
to $A_{x}=2\pi/eL_x$. After the large gauge transformation, 
the gauge potential $A_x=2\pi/eL_x$ will give rise to an 
Aharanov-Bohm phase $e^{-2\pi i p/q}$ when we apply 
$\tau_i$. Therefore, we obtain
\begin{eqnarray}
U_x\tau_i=e^{-2\pi ip/q}\tau_i U_x. 
\label{eq:ux1}
\end{eqnarray}
On the other hand, because $\sigma_i$ and $\rho_i$ do not 
encircle the flux, we have
\begin{eqnarray}
U_x\rho_i =\rho_i U_x,
\quad
U_x\sigma_i =\sigma_i U_x.
\label{eq:ux2}
\end{eqnarray}
Similarly using the adiabatic flux insertion, we can define 
another large gauge transformation $U_y$, in which the 
$y$-component of the gauge potential changes from $A_y=0$ to 
$A_y=2\pi/eL_y$; and we have 
\begin{eqnarray}
U_y \tau_i =\tau_i U_y,
\quad
U_y \rho_i =e^{-2\pi i p/q}\rho_i U_y, 
\quad
U_y \sigma_i =\sigma_i U_y. 
\label{eq:uy}
\end{eqnarray}
\begin{figure}[h]
\begin{center}
\includegraphics[width=3cm]{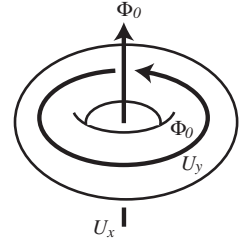}
\caption{Two possible insertions of a unit flux $\Phi_0$.}
\label{fig:flux}
\end{center}
\end{figure}

We notice that the relations (\ref{eq:ux1}), (\ref{eq:ux2}) 
and (\ref{eq:uy}) are compatible with the braid group algebra 
(\ref{eq:braid1}) and (\ref{eq:braid2}). Thus the large gauge 
transformations $U_a$ ($a=x,y$) are (outer) automorphism 
of the braid group operators:
\begin{eqnarray}
\sigma'_i=U_a\sigma_i U_a^{-1},
\quad
\tau'_i=U_a\tau_i U_a^{-1},
\quad
\rho'_i=U_a\rho_i U_a^{-1}.
\end{eqnarray}  
Namely by using relations (\ref{eq:ux1}), (\ref{eq:ux2}) and 
(\ref{eq:uy}), one can check that the new operators $\sigma_i'$, 
$\tau'_i$ and $\rho'_i$ also satisfy the same braid group 
algebra as $\sigma_i$, $\tau_i$ and $\rho_i$.

It is easy to verify that $U_xU_yU_x^{-1}U_y^{-1}$ 
commutes with all $\sigma_i$, $\tau_i$ and $\rho_i$. 
Therefore, by Schur's lemma, for any irreducible 
representation, $U_xU_yU_x^{-1}U_y^{-1}$ is a 
(unimodular) c-number; namely
\begin{eqnarray}
U_xU_y=e^{2\pi i\lambda} U_y U_x.
\label{eq:uxuy}
\end{eqnarray}
As we will see, $\lambda$ is rational and can be fixed
by the requirement of a finite minimal ground state 
degeneracy. It is a new many-body quantum number, also 
characterizing the topological order of the system, 
and is closely related to the fractional quantization 
of Hall conductance (see below).   

Assume that the quasiparticles are Abelian anyons:
\begin{eqnarray}
\sigma_j=e^{i\theta}{\bm 1}, 
\end{eqnarray}
where ${\bm 1}$ is the unit matrix. Then the braid 
group representation is uniquely determined as 
\begin{eqnarray}
\tau_j=e^{-2i\theta(j-1)}T_x,
\quad 
\rho_j=e^{2 i\theta(j-1)}T_y, 
\label{eq:rep_tau_rho}
\end{eqnarray}
with $T_x$ and $T_y$ matrices satisfying
\begin{eqnarray}
T_x T_y=e^{-2i\theta}T_y T_x.
\label{eq:exchangeT}
\end{eqnarray}
On a torus we also have the constraint on $N$ and $\theta$
\begin{eqnarray}
e^{2iN\theta}=1.
\end{eqnarray}
Assuming $N\ge 2$, $\theta/\pi$ must be a rational number, 
$\theta=\pi m/n$, where $m$ and $n$ are mutually prime 
integers. Thus, $T_x$ and $T_y$ satisfy 
\begin{eqnarray}
T_x T_y=e^{-2\pi im/n}T_y T_x. 
\label{eq:exchangeT2}
\end{eqnarray}
The linear automorphisms induced by $U_x$ and $U_y$ now 
reduces to 
\begin{eqnarray}
&&U_x T_x U_x^{-1}=e^{-2\pi i p/q}T_x,
\quad 
U_y T_x U_y^{-1}=T_x,  
\nonumber\\
&&U_x T_y U_x^{-1}=T_y,
\quad
U_y T_y U_y^{-1}=e^{-2\pi i p/q}T_y.
\label{eq:tu}
\end{eqnarray}

To count the ground state degeneracy, we consider the following 
process. First create $N$ pairs of quasiparticle and quasiholes 
out of the ground state, then move the $i$-th quasiparticle by 
$\tau_i$. After it returns to the original position, we pair 
annihilate all quasiparticles and quasiholes. This process 
defines an operation of $\tau_i$ to the ground states. 
Similarly, we define the operation of $\rho_i$ and $\sigma_i$ 
to the ground states. Throughout this letter, we assume that 
Fermi level lies in a gap and the gap remains finite in the 
operations above. Since for a system with Abelian anyonic 
excitations, the operations $\sigma_i$'s on the ground state 
generate merely a phase. Thus we concentrate on the operations 
$\tau_1=T_x$ and $\rho_1=T_y$ on the ground state; from 
Eq. (\ref{eq:rep_tau_rho}) all other $\tau_i$ and $\rho_i$ 
can be expressed in terms of them.

First, we reproduce the degeneracy due to the fractional 
statistics \cite{Birman, Einarsson}. Let us take the basis 
of the ground state to be an eigenstate of $T_x$,
\begin{eqnarray}
T_x |\eta\rangle =e^{i\eta}|\eta\rangle. 
\end{eqnarray}
By applying $T_y$ to $|\eta\rangle$ and using 
(\ref{eq:exchangeT2}), the following new states are obtained: 
\begin{eqnarray}
T_x \left(
T_y^{s}|\eta\rangle 
\right)
=e^{i(\eta-2\pi sm/n)}T_y^{s}|\eta\rangle,
\end{eqnarray}
where $s$ is an integer. Since the new states have $n$ 
different eigenvalues of $T_x$, the ground state has $n$-fold 
degeneracy at least. 

The charge fractionalization gives another constraint for 
the degeneracy. From Eq.(\ref{eq:exchangeT2}), 
$T_x$ and $T_y^n$ commute with each other,
\begin{eqnarray}
T_x T_y^n=T_y^n T_x. 
\end{eqnarray}
Therefore, we can take the basis of the ground states which 
diagonalize $T_x$ and $T_y^n$ simultaneously, 
\begin{eqnarray}
T_x|\eta_1, \eta_2\rangle
=e^{i \eta_1} |\eta_1,\eta_2\rangle,
\quad  
T_y^n|\eta_1, \eta_2\rangle
=e^{i \eta_2} |\eta_1,\eta_2\rangle.
\end{eqnarray}
By applying $U_x$ and $U_y$ to this and using Eq.(\ref{eq:tu}), 
we have
\begin{eqnarray}
&&T_x \left(
U_x^{s}U_y^{t}|\eta_1,\eta_2\rangle
\right)
=e^{i(\eta_1+2\pi sp/q)}U_x^{s}U_y^{t}
|\eta_1,\eta_2\rangle,
\nonumber\\ 
&&T_y^n \left(
U_x^{s}U_y^{t}|\eta_1,\eta_2\rangle \right)
=e^{i(\eta_2+2\pi tnp/q)}U_x^{s}U_y^{t}
|\eta_1,\eta_2\rangle, 
\nonumber\\
\end{eqnarray}
where $s$ and $t$ are integers. If $n/q={\cal N}/{\cal Q}$ 
where ${\cal N}$ and ${\cal Q}$ are mutually prime integers, it is 
found that there are $q{\cal Q}$ sets of eigenvalues of $T_x$ and 
$T_y^n$. This implies that the ground state has $q{\cal Q}$-fold 
degeneracy at least.

By combining the results above, we find that the minimal 
degeneracy of the ground state should be the least common 
multiplet of $n$ and $q{\cal Q}=n{\cal Q}^2/{\cal N}$. 
Namely, the system has {\it $n{\cal Q}^2$-fold} ground 
state degeneracy. This indicates clearly that the fractional 
statistics and the charge fractionalization are {\it both 
responsible} for the ground state degeneracy. The minimal 
degeneracy obtained here includes both the results in 
Ref.\cite{WN} and Ref.\cite{OS} as special cases. More 
possibilities are predicted.  

Up to now, we have not used non-commutating relation 
(\ref{eq:uxuy}) between $U_x$ and $U_y$. This relation
contains an additional parameter $\lambda$, which is 
not fixed uniquely by $e^{*}$ and $\theta$. We believe 
this parameter $\lambda$ could be determined by the 
low-energy effective Lagrangian, which we do not discuss 
here. In order for the degeneracy to be finite, $\lambda$ 
has to be a rational number $\lambda=k/l$, where $k$ and 
$l$ are co-primes. At least in the following examples we 
find that the integers $k$ and $l$ can be determined by 
requiring the degeneracy be minimal given $e^*$ and 
$\theta$. In any case, the degeneracy is given by a 
multiple of $n{\cal Q}^2$.

Now we present some explicit representations of $T_x$, 
$T_y$, $U_x$ and $U_y$ and corresponding degeneracy.   

(1) $\theta=\pi/n$ and $e^{*}=e/n$. -
This corresponds to the Laughlin state with $\nu=1/n$.
Because ${\cal N}={\cal Q}=1$, the minimum degeneracy 
is $n$. If we assume that $U_x$ and $U_y$ satisfy
\begin{eqnarray}
U_x U_y=e^{-2\pi i/n}U_y U_x, 
\end{eqnarray}
we can construct $T_x$, $T_y$, $U_x$ and $U_y$ without 
increasing the degeneracy,
\begin{eqnarray}
&&T_x=S_{n\times n},
\quad
T_y=R_{n\times n},
\nonumber\\
&&U_x=R^{-1}_{n\times n},
\quad
U_y=S_{n\times n}. 
\end{eqnarray}
Here $S_{n\times n}={\rm diag}\{1,e^{i2\pi/n},\cdots,
e^{i2\pi(n-1)/n}\}$ and
\begin{eqnarray}
R_{n\times n}=
\left(
\begin{array}{ccccc}
0&1& 0& \cdots & 0\\
\vdots &0&\ddots&\ddots&\vdots\\
\vdots &\vdots &\ddots &\ddots & 0\\
0&\cdots  &\cdots &0 &1 \\
1&0&\cdots&\cdots&0
\end{array} 
\right).
\end{eqnarray}
They satisfy $S_{n\times n}R_{n\times n}
=e^{-2\pi i/n}R_{n\times n}S_{n \times n}$. This result 
reproduces the degeneracy given in Ref.\cite{WN}.

(2) $\theta=0$ or $\theta=\pi$. -
The quasiparticles are bosons or fermions.
Since $n=1$, we obtain ${\cal N}=1$ and ${\cal Q}=q$. 
Thus the minimal degeneracy is $q^{2}$ \cite{OS}.  
We find that if $U_x$ and $U_y$ commutes with each other,
the minimal degeneracy is realized as
\begin{eqnarray}
&&T_x=R_{q\times q}\otimes 1_{q\times q},
\quad
T_y=1_{q\times q}\otimes R_{q\times q}, 
\nonumber\\
&&U_x=S_{q\times q}^{p}\otimes 1_{q\times q}, 
\quad 
U_y=1_{q\times q}\otimes S_{q\times q}^{p}.
\end{eqnarray}

(3) $q$ and $n$ are mutually prime. -
The degeneracy is $n q^{2}$. We can construct the following 
representation for $T_x$, $T_y$, $U_x$ and $U_y$: 
\begin{eqnarray}
T_x=
1_{q\times q}\otimes R_{q\times q}\otimes S^{m}_{n\times n},
\nonumber\\ 
T_y=
R_{q\times q}\otimes 1_{q\times q}\otimes R_{n\times n},
\nonumber\\
U_x=1_{q\times q}\otimes S_{q\times q}^{p}\otimes 1_{n\times n},
\nonumber\\
U_y=S_{q\times q}^{p}\otimes 1_{q\times q}\otimes 1_{n\times n},
\end{eqnarray}
where the minimal degeneracy is realized. $U_x$ and $U_y$ 
commute with each other in this representation.

(4) $n={\cal N}q$ and $m=1$. -
Because of ${\cal Q}=1$, the minimum degeneracy is $n$.
A representation is given by
\begin{eqnarray}
&&T_x=S_{n\times n},
\quad
T_y=R_{n\times n}, 
\nonumber\\
&&U_x=R_{n\times n}^{-{\cal N}p}, 
\quad 
U_y=S_{n\times n}^{{\cal N}p}.
\end{eqnarray}
$U_x$ and $U_y$ satisfy
$U_x U_y=e^{-2\pi i ({\cal N}p^2/q)}U_y U_x$.

(5) $q={\cal Q}n$. - 
In this case, ${\cal N}=1$, thus the least degeneracy is 
$n{\cal Q}^2$. When ${\cal Q}$ and $n$ are mutually prime 
and $p=m=1$, we can construct the following representation:
\begin{eqnarray}
&&T_x=S_{{\cal Q}\times {\cal Q}}\otimes S_{n\times n}
\otimes 1_{{\cal Q}\times {\cal Q}},
\nonumber\\
&&T_y=1_{{\cal Q}\times {\cal Q}}\otimes R_{n\times n}
\otimes S_{{\cal Q}\times {\cal Q}}, 
\nonumber\\
&&U_x=R_{{\cal Q}\times {\cal Q}}^{-l}\otimes R_{n\times n}^{-k}
\otimes 1_{{\cal Q}\times {\cal Q}}, 
\nonumber\\
&&U_y=1_{{\cal Q}\times {\cal Q}}\otimes S_{n\times n}^{k}
\otimes R_{{\cal Q}\times {\cal Q}}^{-l},
\end{eqnarray}
where $k/n+l/{\cal Q}=1/{\cal Q}n \, ({\rm mod.} 1)$.
In this case, $U_x U_y =e^{-2\pi i k^2/n} U_y U_x$. 

Here we would like to mention that the noncommutativity of the 
large gauge transformation $U_xU_y=e^{2\pi i k/l}U_y U_x$ is 
closely related to the topological order in the fractional 
quantum Hall effect. To see this, consider the degenerate 
ground states $\phi_K$ $(K=1,\cdots,d)$, on a torus with 
boundary conditions parametrized by twisted phases 
$\theta$ and $\varphi$ \cite{NTW}, satisfying 
$U_x|\theta,\varphi \rangle_K =|\theta+2\pi,\varphi \rangle_K$ 
and 
$U_y|\theta,\varphi \rangle_K =|\theta,\varphi+2\pi \rangle_K$.
The Hall conductance is
\begin{eqnarray}
\frac{e^2}{h d} \sum_{K=1}^{d}\int_0^{2\pi}
\int_0^{2\pi} \frac {d\theta d\varphi}{2\pi i}
\left[\left\langle \frac{\partial\phi_K}{\partial\varphi}|
\frac{\partial\phi_K}{\partial \theta}\right\rangle
-(\theta\leftrightarrow\varphi) \right].\nonumber
\end{eqnarray} 
Because of $U_x^{l}U_y=U_y U_x^l$, we can take the basis which
diagonalizes both $U_x^{l}$ and $U_y$. In this basis a change 
in $\theta$ by $2\pi l$ or in $\varphi$ by $2\pi$ leads the 
state back to itself. Therefore, we have a torus with
$0\le\theta <2\pi l$ and $0\le\varphi <2\pi$. The above 
integral can be recast into
\begin{eqnarray}
\frac{e^2}{h d} \sum_{r=0}^{d/l-1}
\int_0^{2\pi l}\int_0^{2\pi} \frac{d\theta d\varphi}{2\pi i}
\left[\left\langle \frac{\partial\phi_{rl+1}}{\partial\varphi}|
\frac{\partial\phi_{rl+1}}{\partial \theta}\right\rangle
-(\theta\leftrightarrow\varphi) \right], 
\nonumber
\end{eqnarray}
since the degenerate ground states $\phi_K$ satisfy 
$\phi_{rl+m}(\theta+2\pi,\varphi)=\phi_{rl+m+1}(\theta,\varphi)$
$(r=0,\cdots, d/l-1, m=1,\cdots l-1)$.
Therefore we have the following formula for the Hall 
conductance, generalizing the result of Ref.\cite{NTW}: 
\begin{eqnarray} 
\sigma_{xy}=\frac{e^2}{hd}\sum_{r=0}^{d/l-1}I_r
=\frac{e^2}{h}\frac{I}{l}.
\end{eqnarray}
Here $I_r$ is a generalized TKNN integer \cite{TKNN} defined 
by $\phi_{rl+1}$. For the second equality, we have noted  
that all $I_r$'s take the same value $I$, since the 
degenerate ground states are related to each other by 
symmetry operations $T_x$ and $T_y$. This indicates clearly 
that the noncommutativity $\lambda=k/l$ of the large gauge 
transformations is essential to fractional quantization of 
the Hall conductance.

The fractional charge we discussed above can be any conserved 
$U(1)$ quantum number, with the flux threading understood as 
twisted boundary conditions. The minimal degeneracy obtained 
can be generalized to a high-genus Riemann surface: If 
the genus is $g$, for Abelian topological orders we find $g$ commuting
copies of the discrete algebra presented above, so the minimal degeneracy is 
$n^{g}{\cal Q}^{2g}$.

To conclude, in this letter we have proposed a discrete 
symmetry algebra, Eqs.(\ref{eq:uxuy}), (\ref{eq:exchangeT2}) and 
(\ref{eq:tu}), of the operations $T_x$, $T_y$, $U_x$ and $U_y$, 
that completely characterizes the ground state subspace of a 
generic {\it Abelian} topological order, i.e. that supports
Abelian anyonic excitations, on a torus. The identification 
verifies the old idea that the ground state degeneracy in a 
topological phase is due to the emergence of a discrete 
symmetry \cite{TW,WHK}. We note that the algebra identified is indeed 
of topological origin and contains only three fractional 
parameters: quasiparticle charge $e^*/e$, anyon statistics 
$\theta/2\pi$ and flux noncommutativity $\lambda$. Ground 
state degeneracy is determined by the representations of 
this symmetry algebra.  

{\it Acknowledgement} The work was supported in part by the U.S. NSF
through grant No. PHY-0407187 (YSW). This work was begun in summer
2005, when YSW visited the ISSP, University of Tokyo. He thanks the 
financial support and warm hospitality from the host institution.   


\bibliography{SKW}

\end{document}